\documentclass{aa}
\usepackage{graphicx}
\begin{document}
   \title{Detection of DCO$^+$ in a circumstellar disk}


   \author{Ewine F.\ van Dishoeck
          \inst{1}
          \and
          Wing-Fai Thi\inst{1,2}
          \and
          Gerd-Jan van Zadelhoff\inst{1}
          }

   \offprints{E.F.\ van Dishoeck}

   \institute{Leiden Observatory, P.O. Box 9513, 2300 RA Leiden, 
              The Netherlands \\
              \email{ewine@strw.leidenuniv.nl}
         \and
             Department of Physics \& Astronomy, University College
            London, Gower Street, London WC1E 6BT, U.K.\\
             }

   \date{Received December 5, 2002; Accepted January 22, 2003}

   \abstract{
We report the first detection of DCO$^+$ in a circumstellar disk.
The DCO$^+$ $J$=5--4 line at 360.169 GHz is observed with the 15m
James Clerk Maxwell Telescope in the disk around the pre-main sequence
star TW Hya.  Together with measurements of the HCO$^+$ and
H$^{13}$CO$^+$ $J$=4--3 lines, this allows an accurate determination
of the DCO$^+$/HCO$^+$ ratio in this disk. The inferred value of
0.035$\pm$0.015 is close to that found in cold pre-stellar cores and
is somewhat higher than that measured in the envelope around the
low-mass protostar IRAS 16293 -2422. It is also close to the DCN/HCN
ratio obtained for pristine cometary material in the jet of comet
Hale-Bopp.  The observed DCO$^+$/HCO$^+$ ratio for TW Hya is
consistent with theoretical models of disks which consider gas-phase
fractionation processes within a realistic 2-D temperature
distribution and which include the effects of freeze-out onto grains.
               }



   \maketitle
%

\section{Introduction}

Disks around pre-main sequence stars are the likely sites of the
formation of planetary systems (e.g., Beckwith 1999).  Part of the gas
and dust from the collapsing protostellar envelope settles into this
disk, where it may undergo a complex chemistry before incorporation
into planets and icy bodies.  Tracing this chemistry is a key goal of
molecular astrophysics, not only to establish a chemical inventory
prior to planet formation but also as a probe of the dynamical
processes in disks such as radial and vertical mixing (e.g., Aikawa et
al.\ 1999, Markwick et al.\ 2002).  Moreover, the abundances and
excitation of the molecules provide unique insight into the physical
structure of disks such as their temperature and density profiles
(Dutrey et al.\ 1997, van Zadelhoff et al.\ 2001).

Deuterated molecules are a particularly interesting probe of the
temperature history of interstellar and circumstellar gas. It is well
known that the abundances of deuterated molecules in cold pre-stellar
cores and protostellar envelopes are enhanced by orders of magnitude
over the elemental [D]/[H] abundance ratio of $1.5\times 10^{-5}$
through fractionation (e.g., Watson 1976, Millar et al.\ 1989).  In
particular, at temperatures below $\sim$50 K, the reaction
$$ \rm H_3^+ \ + \ HD 
   {{{\null}\atop \rightarrow} \atop {\leftarrow \atop {\null}}} 
   \rm  H_2D^+ \ +\  H_2  \ + \ 232 \ K \eqno(1)$$
is driven strongly in the forward direction (e.g., Pagani et al.\
1992). H$_2$D$^+$ can subsequently transfer a deuteron to the abundant
CO molecule
$$ {\rm H_2D^+ \ + \ CO \to \ DCO^+ \ + \ H_2 } \eqno(2)$$
leading to DCO$^+$/HCO$^+$ ratios of order 0.01. Such high ratios have
indeed been observed in interstellar clouds (e.g., Gu\'elin et al.\
1977, Wootten et al.\ 1982, Butner et al.\ 1995, Williams et al.\
1998) and in protostellar regions (van Dishoeck et al.\ 1995, Loinard
et al.\ 2000, Shah \& Wootten 2001, Parise et al.\ 2002).  Deuterium
fractionation can also occur through reactions of atomic deuterium on
the surfaces of interstellar grains (Tielens 1983), but this is
thought to affect primarily molecules which are formed on grains such
as H$_2$CO and CH$_3$OH. The DCO$^+$/HCO$^+$ ratio can thus serve as a
clear measure of the importance of low-temperature gas-phase deuterium
fractionation processes. Moreover, this ratio is enhanced if the main
destroyer of H$_3^+$, ---i.e., CO---, is depleted onto grains, since
reaction (1) then becomes an important H$_3^+$ removal route
in addition to recombination with electrons,
enhancing H$_2$D$^+$ even more (Brown \& Millar 1989, Stark et al.\
1999, Caselli et al.\ 1999). Thus, the DCO$^+$/HCO$^+$ ratio can also
trace the level of depletion.

We present here the first detection of the DCO$^+$ ion in a
circumstellar disk, using the James Clerk Maxwell Telescope
(JCMT). Together with observations of the HCO$^+$ ion and its
optically thin isotope H$^{13}$CO$^+$, an accurate DCO$^+$/HCO$^+$
ratio is derived.  This ratio is subsequently compared with that found
in the envelopes of deeply embedded protostars --- the precursor
material of disks ---, and with those observed in solar system
objects, in particular comets --- the remnant material of disks.

\section{Observations}

The DCO$^+$ $J$=5--4 rotational line at 360.169 GHz was observed with
the JCMT\footnote{The James Clerk Maxwell Telescope is operated by
the Joint Astronomy Centre in Hilo, Hawaii, on behalf of the Particle
Physics and Astronomy Research Council in the United Kingdom, the
National Research Council of Canada and the Netherlands Organization
for Scientific Research.} on Mauna Kea, Hawaii. The dual polarization
receiver B3 was used in single-sideband mode. The backend was the
Digital Autocorrelator Spectrometer (DAS) set at a resolution of
$\sim$0.15 km s$^{-1}$. The observations were taken in beam-switching
mode with a throw of 180$''$. The same set-up was used to observe the
HCO$^+$ and H$^{13}$CO$^+$ $J$=4--3 lines at 356.734 and 346.998 GHz,
respectively.  The antenna temperatures were converted to main-beam
temperatures using a beam efficiency of 62\%, calibrated from
observations of planets by the telescope staff. The flux calibration
was checked against bright sources with well-determined fluxes for
different lines and was generally found to agree within 10\%.
The data were reduced and analyzed in the SPECX and CLASS data
reduction packages.

The DCO$^+$ search focussed on the disk around TW Hya, an isolated T
Tauri star (K8Ve) which is part of a young group of stars at a
distance of only $\sim$56 pc (Webb et al.\ 1999).  Its age is
estimated to be $\sim$7--15 Myr, somewhat older than most classical T
Tauri stars, but its large lithium abundance and H$\alpha$ equivalent
width are indicative of continued active disk accretion at a rate of
$10^{-8}$ M$_{\odot}$ yr$^{-1}$ (Kastner et al.\ 2002). TW Hya is
surrounded by a nearly face-on disk of $\sim 2\times 10^{-2}$ M$_{\odot}$
(Wilner et al.\ 2000), which has been imaged recently in scattered
light with the Hubble Space Telescope (Krist et al.\ 2000). The extent
of the disk in the scattered light is $\sim$200 AU, corresponding to a 3.6$''$
radius at 56 pc. Since this size is small
compared with the $\sim$13$''$ beam size of the JCMT at
360 GHz, the data suffer from beam dilution. Thus, long
integration times of typically 2--4 hrs were needed to detect the weak
lines reported here.

   \begin{figure}
   \centering
   \includegraphics[width=7cm,angle=-90]{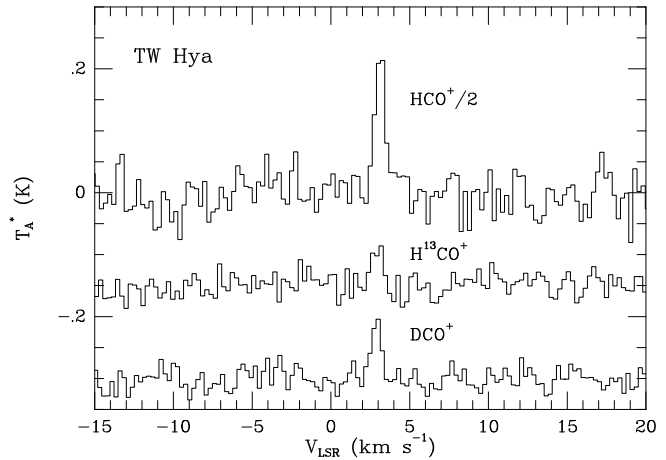}
      \caption{
JCMT observations of the HCO$^+$ $J$=4--3, H$^{13}$CO$^+$ $J$=4--3
and DCO$^+$ $J$=5--4 lines toward TW Hya. The H$^{13}$CO$^+$
and DCO$^+$ spectra have been shifted by --0.15 and --0.3 K, respectively.
              }
         \label{fig1}
   \end{figure}

The DCO$^+$ and HCO$^+$ observations are part of a larger survey for
molecular lines from disks, which will be reported in Thi et al.\
(2003).

\section{Results}

In Figure~1, the HCO$^+$, H$^{13}$CO$^+$ and DCO$^+$ spectra toward TW
Hya are presented. All three lines are clearly detected at the same
velocity, $V_{\rm LSR}$=3.0$\pm$0.1 km s$^{-1}$. The lines are very narrow,
$\Delta V \approx 0.6$ km s$^{-1}$, and do not show the characteristic
double-peak profile of a rotating disk, consistent with the nearly
face-on orientation. The lack of detection of the $^{12}$CO 3--2 line
at 30$''$ offset positions demonstrates that the molecular emission is
indeed associated with the disk and not due to any remnant cloud
material in the vicinity (Thi et al.\ 2003).  The critical densities
of the observed HCO$^+$ and DCO$^+$ lines are at least $10^6$
cm$^{-3}$, further assuring that only the dense gas in the disk is probed.
The integrated line intensities are summarized in Table~1.

\begin{table} %
\caption[]{Summary of observational data\label{tab1}}
\begin{tabular}{llll}                                        
\hline
Species & Line &$\int T_{\rm MB} dV$ & $N_{200}^a$ \\
    &  & (K km s$^{-1}$) & ($10^{12}$ cm$^{-2}$) \\
\hline
HCO$^+$ & $J$=4--3 & 0.49 & 8.5$^b$\\
H$^{13}$CO$^+$ & $J$=4--3 & 0.07 & 0.14 \\
DCO$^+$ & $J$=5--4 & 0.11 & 0.30\\
\hline
\end{tabular}
\begin{flushleft}
$^a$ Column density derived under the assumption of LTE excitation at 25~K
for a 200 AU radius disk. \\

$^b$The HCO$^+$ value is obtained from the optically
thin H$^{13}$CO$^+$ column density assuming [$^{12}$C]/[$^{13}$C]=60.
\end{flushleft}
\end{table}

Molecular column densities and abundances can be derived from the
observed line strengths under different assumptions. In general, the
emerging lines from interstellar gas are a complex function of
abundance, excitation and radiative transfer effects. In circumstellar
disks, where strong chemical gradients can exist in both vertical and
radial directions, the interpretation of spatially unresolved data is
particularly uncertain. However, since the main interest of this study
is in the DCO$^+$/HCO$^+$ ratio, the analysis can be considerably
simplified if it is assumed that DCO$^+$ and HCO$^+$ occupy roughly the same
regions of the disk. Moreover, since the densities in disks are
generally high in the regions where these molecules exist ($>10^6$
cm$^{-3}$) (van Zadelhoff et al.\ 2001), the excitation can be assumed
to be in local thermal equilibrium to first order. A single excitation
temperature of 25 K is adopted in the following analysis.  Finally, both the
H$^{13}$CO$^+$ and DCO$^+$ lines are assumed to be optically thin. The
observed line intensity ratio of HCO$^+$/H$^{13}$CO$^+$ of $\sim$7 compared
with the optically thin ratio of $\sim$60 clearly indicates that the main
isotope HCO$^+$ $J$=4--3 line is optically thick.

The inferred column densities for a 200 AU radius disk are included in
Table 1.  For an assumed  gas + dust disk mass of $2\times
10^{-2}$ M$_{\odot}$, the beam-averaged HCO$^+$ abundance is $\sim
2\times 10^{-11}$, derived from the optically thin H$^{13}$CO$^+$
line assuming LTE excitation at 25 K (see Thi et al.\ 2003 for
details).  This is significantly lower than the typical HCO$^+$
abundance of $\sim 10^{-9}-10^{-8}$ in dark clouds (e.g., Ohishi et
al.\ 1992) and some protostellar envelopes (e.g., Sch\"oier et al.\
2002). Most of the observed emission is thought to originate from the
warm intermediate layer of the disk just below the surface where the
HCO$^+$ abundance reaches a few $\times 10^{-10}$ (e.g., Aikawa et
al.\ 2002, Willacy \& Langer 2000).  The molecules are significantly
depleted in the cold midplane of the disk where the bulk of the mass
resides .  For the TW Hya disk, observations of CO and other species
indicate depletions up to a factor of 100 in the midplane (van
Zadelhoff et al.\ 2001).

\section{Discussion}

The inferred column densities imply a beam-averaged DCO$^+$/HCO$^+$
abundance ratio of 0.035 $\pm$ 0.015, assuming a [$^{12}$C]/[$^{13}$C]
isotopic ratio of 60 (see Table~1). The error bar reflects the
observational uncertainties. The DCO$^+$/HCO$^+$ value is more than
three orders of magnitude higher than the elemental [D]/[H] ratio of
$1.5\times 10^{-5}$, illustrating that strong deuterium fractionation
occurs in disks.

Models of the deuterium fractionation in disks have been calculated by
Aikawa \& Herbst (1999, 2001) and Aikawa et al.\ (2002), with the
latter models using a realistic 2-dimensional (2D) temperature and
density structure of a flaring disk. The models include a detailed
gas-phase chemistry network with freeze-out onto grains, but do not
contain an active grain-surface chemistry. The resulting
DCO$^+$/HCO$^+$ abundance is found to decrease with decreasing radius
from $\sim 0.1$ at 400 AU to $<0.01$ at $<$50 AU, owing to the
increasing temperature in the inner disk.  The DCO$^+$/HCO$^+$ ratio
also decreases with height in the outer disk, because of the higher
temperatures in the upper layers.  Close to the midplane where strong
CO freeze-out occurs, the DCO$^+$/HCO$^+$ ratio reaches high values,
but this region contributes negligibly to the observed emission due to
the much lower overall abundances.  Time-dependent effects appear to
play a minor role, although results for disks as old as 10 Myr have
not been published.  Overall, the observed DCO$^+$/HCO$^+$ ratio of
0.035$\pm$0.015 averaged over the entire TW Hya disk appears
consistent with these models.

Roberts et al.\ (2002) show that the model results for molecular
clouds are lowered by a factor of a few if updated rate coefficients
for reaction (1) are used. Since the main processes affecting the
DCO$^+$/HCO$^+$ ratio in disks, namely low-temperature gas-phase
reactions and freeze-out of CO on grains, are the same as those in
clouds, similar effects are expected for disk models using updated
rate coefficients.

\begin{table*} %
\caption[]{Deuterium fractionation ratios in 
different environments \label{tab2}}
\begin{tabular}{lllll}                                        
\hline
Type of region & Object & Species & D/H & Reference \\
\hline
Dark cores & Various  & DCO$^+$ & 0.02--0.07 & Butner et al.\ (1995) \\  
           & Various  & DCO$^+$ & 0.02--0.06 & Williams et al.\ (1998) \\
           & TMC-1 CP-peak & DCO$^+$ & 0.012 & Turner (2001) \\
           & L1544 & DCO$^+$ & 0.04 & Caselli et al.\ (2002) \\
           & L1689N D-peak & DCO$^+$ & 0.08 & Lis et al.\ (2002) \\
           & L134N & DCO$^+$ & 0.18 & Tin\'e et al.\ (2000) \\
Low-mass     & IRAS 16293-2422 & DCO$^+$ & 0.009 
                                           & Sch\"oier et al.\ (2002)\\
protostars      & &DCN & 0.012 & Sch\"oier et al.\ (2002) \\
                   & NGC 1333 I4A & DCO$^+$ & 0.01 & Stark et al.\ (1999) \\
                 & Various & DCO$^+$ & 0.005--0.035 & Shah \& Wootten (2001) \\
                   &         & DCN  & 0.01--0.02 & Shah \& Wootten (2001) \\
Disk           & TW Hya & DCO$^+$ & 0.035 & This work \\
Comet         & Hale-Bopp jet & DCN  & 0.09 & Blake et al.\ (1999) \\
             & Hale-Bopp coma & DCN & 0.002 & Meier et al.\ (1998) \\
Disk model   &  Outer $\geq$200 AU  & DCO$^+$ & 0.05 & Aikawa et al.\ (2002)\\
             &          & DCN & 0.03 & Aikawa et al.\ (2002)\\
             & Inner 50 AU   & DCO$^+$ & 0.002 &Aikawa et al.\ (2002)\\
             &               & DCN & 0.001 &Aikawa et al.\ (2002)\\
\hline
\end{tabular}
\begin{flushleft}
\end{flushleft}
\end{table*}

Only few other measurements of deuterated molecules in disks have
been reported. Qi (2001) and Kessler et al.\ (2003) searched for DCN
and HDO in several disks using the Owens Valley Millimeter Array
but obtained mostly upper limits.  The inferred DCN/HCN ratio of
$<0.002$ is much lower than that for DCO$^+$/HCO$^+$ found here.

Table 2 summarizes the observed DCO$^+$/HCO$^+$ and DCN/HCN ratios in
different environments. It is seen that the DCO$^+$/HCO$^+$ ratio in
the TW Hya disk is comparable to that found in cold dark cores where
freeze-out has been observed (e.g., L1544), testifying to the low
temperatures in disks.  Indeed, DCO$^+$/HCO$^+$ ratios as high as
0.035 are difficult to produce in models which do not include CO
depletion (Roberts et al.\ 2002).  It is higher than that found in
most protostellar envelopes, where heating has affected a larger
fraction of the material resulting in higher temperatures and less CO
freeze-out.

Because comets spend much of the time since their formation in the
cold outer region of the Solar System, they are likely to contain the
most primitive record of solar nebula material.  No DCO$^+$ 
is observed in comets, but DCN/HCN has been measured. In cold clouds
and protostellar envelopes, the observed DCN/ HCN ratios are often
comparable to the DCO$^+$/HCO$^+$ ratios (see Table~2), even though
they involve different fractionation reactions. The most pristine
material originating from below the comet surface and emanating in
jets shows very high DCN/HCN ratios comparable to those seen in cold
dark clouds and in the TW Hya disk.  The upper surface layers which
evaporate to produce the coma have significantly lower ratios,
indicative of processing in the solar nebula.

The similarity of the deuterium fractionation ratios in cold clouds,
disks and pristine cometary material suggests that the gas spends most
of its lifetime at low temperatures and is incorporated into the disks
before the envelope is heated, i.e., before the Class I
stage. Alternatively, the DCO$^+$/HCO$^+$ ratio may be reset in disks
by low-temperature gas-phase chemistry. Comparison of D/H ratios of
molecules which enter the disk in the gas phase (such as HCO$^+$) and
those which are likely incorporated as ices (e.g., H$_2$CO, CH$_3$OH)
can distinguish between these scenarios. Also, spatially resolved data
can further test the models, since the high ratios observed here probe
largely the outer disk.

The data presented in this paper are at the limit of the capabilities of
current observational facilities.  The Atacama Large Millimeter Array
(ALMA) will be essential to provide high-resolution measurements of
D/H ratios in disks down to comet- and planet-forming regions.

\begin{acknowledgements}
Astrochemistry in Leiden is supported by a Spinoza
grant from the Netherlands Organization of Scientific Research (NWO).
WFT thanks PPARC for a post-doctoral grant to UCL.

\end{acknowledgements}

\end{document}